\documentclass[11pt]{article}
\newcommand{\Comment}[1]{{}}
\usepackage{amsfonts,amsthm,amsmath,amssymb,slashed}
\usepackage[textwidth = 430 pt, textheight = 630 pt]{geometry}
\usepackage{color}

\Comment{\usepackage{color}
\definecolor{MyDarkBlue}{rgb}{0.15,0.15,0.45}
\usepackage[linktocpage=true]{hyperref}
\hypersetup{
colorlinks=true,
citecolor=MyDarkBlue,
linkcolor=MyDarkBlue,
urlcolor=MyDarkBlue,
pdfauthor={Heliudson Bernardo and Horatiu Nastase},
pdftitle={The title},
pdfsubject={hep-th}
}

\usepackage[numbers,sort&compress]{natbib}
\usepackage{hypernat}}
\usepackage{graphicx}
\usepackage{cite}

\newcommand\ignore[1]{}
\def\one{{\,\hbox{1\kern-.8mm l}}}

\def\Tr{{\rm Tr\, }}

\def\d{\partial}

\def\Tr{\mathop{\rm Tr}\nolimits}

\newcommand{\Cset}{{\,\,{{{^{_{\pmb{\mid}}}}\kern-.45em{\mathrm C}}}}}

\newcommand{\be}{\begin{equation}}
\newcommand{\bea}{\begin{eqnarray}}

\newcommand{\ee}{\end{equation}}
\newcommand{\eea}{\end{eqnarray}}

\begin{document}

\renewcommand{\thefootnote}{\fnsymbol{footnote}}

\makeatletter
\@addtoreset{equation}{section}
\makeatother
\renewcommand{\theequation}{\thesection.\arabic{equation}}

\rightline{}
\rightline{}




\begin{center}
{\LARGE \bf{\sc Holographic cosmology from "dimensional reduction" of ${\cal N}=4$ SYM vs. $AdS_5\times S^5$}}
\end{center} 
 \vspace{1truecm}
\thispagestyle{empty} \centerline{
{\large \bf {\sc Heliudson Bernardo${}^{a}$}}\footnote{E-mail address: \Comment{\href{mailto:heliudson@gmail.com}}{\tt heliudson@gmail.com}}
{\bf{\sc and}}
{\large \bf {\sc Horatiu Nastase${}^{a}$}}\footnote{E-mail address: \Comment{\href{mailto:horatiu.nastase@unesp.br}}{\tt horatiu.nastase@unesp.br}}
}

\vspace{.5cm}


\centerline{{\it ${}^a$Instituto de F\'{i}sica Te\'{o}rica, UNESP-Universidade Estadual Paulista}} 
\centerline{{\it R. Dr. Bento T. Ferraz 271, Bl. II, Sao Paulo 01140-070, SP, Brazil}}

\vspace{1truecm}

\thispagestyle{empty}

\centerline{\sc Abstract}

\vspace{.4truecm}

\begin{center}
\begin{minipage}[c]{380pt}
{\noindent We propose a way to obtain holographic cosmology models for 3+1 dimensional cosmologies vs. 3 dimensional field theories from a "dimensional reduction"
procedure, obtained by integrating over the time direction, of (modifed) standard holographic duals of 3+1 dimensional field theories. The example of a modified ${\cal N}=4$ 
SYM vs. $AdS_5\times S^5$ is presented, and in perturbation theory doesn't match observations, though at strong coupling it might. But the proposed mechanism is 
more general, and it could in principle be applied to other top down holographic models. 

}
\end{minipage}
\end{center}

\vspace{.5cm}

\setcounter{page}{0}
\setcounter{tocdepth}{2}

\newpage

\renewcommand{\thefootnote}{\arabic{footnote}}
\setcounter{footnote}{0}

\linespread{1.1}
\parskip 4pt



\section{Introduction}

The idea of a holographic cosmology has been around for a long time. The first concrete proposal of how that would look like was put forward by Maldacena in 
\cite{Maldacena:2002vr}, stating that the wave function of the Universe, as a function of spatial 3-metrics (and scalars), $\psi[h_{ij},\phi]$ in some gravity dual background
(in his specific case, proposed for some space that asymptotes to de Sitter), equals the partition function of some (3 dimensional) field theory, with sources (for the 
energy-momentum tensor $T_{ij}$ and some scalar operator ${\cal O}$) $h_{ij},\phi$, i.e., $Z[h_{ij},\phi]=\psi[h_{ij},\phi]$. However, at the time, there was no concrete 
proposal for a gravity dual pair. 

In \cite{McFadden:2009fg}, such a model was proposed, and a sort of phenomenological holographic cosmology approach was born. It was first noted that, for 
cosmological scale factors $a(t)$ that are both exponential (as in standard inflation, and corresponding to AdS space) or power law (as in power law inflation, and
corresponding to nonconformal D-branes, for instance), a specific Wick rotation, the "domain wall/cosmology correspondence", turns the cosmology into a 
standard holographic space like a domain wall, that should have a field theory dual in 3 Euclidean dimensions. A holographic computation then relates 
the cosmological power spectrum, coming from the $\langle \delta h_{ij}(\vec{x})\delta h_{kl}(\vec{y})\rangle$ correlators in the bulk, with 
$\langle T_{ij}(\vec{x}) T_{kl}(\vec{y})\rangle $ correlators in the boundary field theory. One can assume a regime where the field theory is perturbative, and the 
latter correlators can be calculated from Feynman diagrams. Then by comparing the cosmological power spectrum with CMBR data, we can find the best fit 
in a phenomenological class of field theories, with a "generalized conformal structure". 
In \cite{Afshordi:2016dvb,Afshordi:2017ihr} (see  \cite{Easther:2011wh} for an early attempt to 
match to the CMBR, in WMAP data)
it was shown that the phenomenological fit matches the CMBR as 
well as the (different) standard $\Lambda$CDM with inflation, though the perturbative field theory approximation breaks down for modes with $l<30$. 
But this holographic cosmology paradigm is more general than the specific class of phenomenological models: it includes standard inflationary cosmology, 
where the gravitational side is weakly coupled, as well as intermediate coupling field theory models, that can be treated non-perturbatively on the lattice.\footnote{Lattice
work on this is ongoing.}

Another approach to holographic cosmology was considered in \cite{Awad:2008jf,Brandenberger:2016egn,Ferreira:2016gfg}, where one starts with a "top down" construction (a well-defined 
gravity dual pair, derived as the decoupling limit of some system of branes), 
specifically a modified version of the original ${\cal N}=4$ SYM vs. string theory in $AdS_5\times S^5$, where an FLRW cosmology with $a(t)$ replaces the Minkowski 
metric, and a nontrivial dilaton is introduced. On the field theory side, one has a time-dependent coupling now. 
The model has been used in \cite{Brandenberger:2016egn,Ferreira:2016gfg} to show how perturbations entering a Big Crunch exit after the Big Bang, one issue
that has been very contentious in ekpyrotic and cyclic cosmologies. It was shown that the spectral index of perturbations exits unchanged, but there was no simple mechanism in \cite{Brandenberger:2016egn,Ferreira:2016gfg} of calculating the 
power spectrum of fluctuations for CMBR.

A natural question to ask then is: can one modify the top down construction of \cite{Awad:2008jf,Brandenberger:2016egn,Ferreira:2016gfg}, to fit it into the 
holographic paradigm of \cite{McFadden:2009fg}, for which the common concrete realization so far is a phenomenological (bottom up) approach?
In this paper, we want to give an answer in the affirmative. We will find that we can modify the general proposal of Maldacena for $Z[h_{ij},\phi]=\psi[h_{ij},\phi]$
to deal with this case of having both time and a radial coordinate, and then use an integration over the time coordinate, from close to zero until an arbitrary  time $t_0$ (but not to the future of it, in this way obtaining a function of $t_0$), to argue 
that we have effectively a "dimensional reduction" over the time direction. The result is a specific theory with "generalized conformal structure", but we will see that 
in perturbation theory it doesn't fit the CMBR data. However, it could be that by considering a  nonperturbative coupling, we have a match. It could also be that 
one has to apply the above procedure to some other top down holographic duality construction. 

The paper is organized as follows. In section 2 we review the holographic cosmology paradigm of McFadden and 
Skenderis. In section 3 we consider the top-down model coming from the ${\cal N}=4$ SYM model vs. $AdS_5\times 
S^5$, and present our proposal for the extension of the Maldacena map, and the resulting ``dimensional reduction'' 
in the time direction. We also show that the dilaton transforms in the bulk, resulting in an operator VEV on the boundary, 
that depends on the cosmological solution. In section 4 we conclude.

\section{Holographic cosmology paradigm}

In this section we review the holographic cosmology paradigm of \cite{McFadden:2009fg}.\footnote{See 
\cite{Larsen:2003pf} for an early attempt to relate inflation with holography} 
One considers a cosmological FLRW model, coupled with a scalar $\phi$, 
and having fluctuations in both, 
\bea
ds^2&=&-dt^2+a^2(t)[\delta_{ij}+h_{ij}(t,\vec{x})]dx^i dx^j\;,\cr
\phi(t,\vec{x})&=&\phi(t)+\delta\phi(t,\vec{x})a.
\eea

After a Wick rotation, the "domain wall/cosmology correspondence", putting $t=-iz$, but also $\bar \kappa^2=-\kappa^2, \bar q =-iq$ (here $\kappa$ is the Newton constant
and $q$ is momentum), which in field theory corresponds to $\bar q=-iq,\bar N=-iN$, we obtain the domain wall gravity dual 
\bea
ds^2&=&+dz^2+a^2(z)[\delta_{ij}+h_{ij}(z,\vec{x})]dx^i dx^j\;,\cr
\phi(z,\vec{x})&=&\phi(z)+\delta\phi(z,\vec{x})a\;,
\eea

The generic "domain wall" above can correspond to (asymptotically) AdS space, for (asymptotically) exponential $a(z)$, in which case expect a field 
theory that is conformal in the UV. Or it can correspond to some holographic dual of the type of nonconformal branes, for power law $a(z)$, in which case one 
expects a "generalized conformal structure": the theory has as only dimensional parameter the YM coupling $g_{YM}$, which appears as an overall 
factor in front of the action. Therefore it is of the type that we would obtain by dimensionally reducing a 4 dimensional conformal field theory. 

Specifically, the phenomenological class of models considered for the fit to the CMBR is a super-renormalizable theory of $SU(N)$ gauge fields $A_i^a$, scalars 
$\phi^{aM}$ and fermions $\psi^{aL}$, where $a$ is an adjoint $SU(N)$ index and $M,L$ are flavour indices, with action
\bea
S_{\rm QFT}&=&\int d^3x \Tr\left[\frac{1}{2}F_{ij}F^{ij}+\delta_{M_1M_2}D_i\Phi^{M_1}D^i\Phi^{M_2}+2\delta_{L_1L_2}\bar \psi^{L_1}\gamma^i D_i
\psi^{L_2}\right.\cr
&&\left.+\sqrt{2}g_{YM}\mu_{ML_1L_2}\Phi^M \bar \psi^{L_1}\psi^{L_2}+\frac{1}{6}g^2_{YM}\lambda_{M_1...M_4}\Phi^{M_1}...\Phi^{M_4}\right]\cr
&=&\frac{1}{g^2_{YM}}\int d^3x \Tr\left[\frac{1}{2}F_{ij}F^{ij}+\delta_{M_1M_2}D_i\Phi^{M_1}D^i\Phi^{M_2}+2\delta_{L_1L_2}\bar \psi^{L_1}\gamma^i D_i
\psi^{L_2}\right.\cr
&&\left.+\sqrt{2}\mu_{ML_1L_2}\Phi^M \bar \psi^{L_1}\psi^{L_2}+\frac{1}{6}\lambda_{M_1...M_4}\Phi^{M_1}...\Phi^{M_4}\right].\label{phenoaction}
\eea
Here $\lambda_{M_1...M_4}$ and $\mu_{ML_1L_2}$ are dimensionless, and only $g_{YM}$ is dimensional, and in the second line the fields have been rescaled 
by $g_{YM}$ in order to obtain $g_{YM}$ as an overall factor, and the dimensions of the fields to be the ones in 4 dimensions.

The generalized conformal structure means that the momentum dependence organizes into a dependence on the effective dimensionless coupling of the theory,
\be
g^2_{\rm eff}=\frac{g^2_{YM}N}{q}.
\ee
Correlators will thus depend on $g^2_{\rm eff}$, and in perturbation theory one obtains, as usual, a combination of powers of $g^2_{\rm eff}$ and 
$\ln g^2_{\rm eff}$.

The CMBR power spectrum is defined in terms of the standard scalar and tensor fluctuations in momentum space $\zeta(q)$ and $\gamma_{ij}(q)$ as 
\bea
 \Delta_S^2(q)&\equiv& \frac{q^3}{2\pi^3}\langle \zeta(q)\zeta(-q)\rangle \cr
 \Delta_T^2(q)&\equiv & \frac{q^3}{2\pi^3}\langle \gamma_{ij}(q)\gamma_{ij}(-q)\rangle.
 \eea

In principle, one could relate them to the two-point functions of the energy-momentum tensors via the Maldacena relation $Z[h_{ij}]=\psi[h_{ij}]$ as follows. 
From general theory, the partition function is represented as the generating functional of correlators as
\be
Z[h_{ij}]=\exp \left[\int \frac{1}{2}h^{ij}\langle T_{ij}T_{kl}\rangle h^{kl}+...\right]\;,
\ee
which leads to the 2-point function of cosmological fluctuations $h_{ij}$ as
\be
\langle h_{ij} h_{kl}\rangle =\int {\cal D} h_{mn} |\psi[h_{pq}]|^2h_{ij}h_{kl}\sim \frac{1}{{\rm Im} \langle T_{ij}T_{kl}\rangle}\;,\label{correl}
\ee
where the last equality is qualitative, and involves a nontrivial calculation.

The more precise relation was found in \cite{McFadden:2010na}, based on the formalism in \cite{Papadimitriou:2004ap,Papadimitriou:2004rz}, and is reviewed in the 
Appendix. Decomposing the energy-momentum tensor correlators as 
\be
\langle T_{ij}(\bar q)T_{kl}(-\bar q)\rangle =A(\bar q) \Pi_{ijkl}+B(\bar q)\pi_{ij}\pi_{kl}\;,
\ee
where 
\be
\Pi_{ijkl}=\pi_{i(k}\pi_{l)j}-\frac{1}{2}\pi_{ij}\pi_{kl}\;,\;\;\;
\pi_{ij}=\delta_{ij}-\frac{\bar q_i\bar q_j}{\bar q^2}
\ee
are the 4-index transverse traceless projection operator ($\Pi_{ijkl}$), and the 2-index transverse projection operator
($\pi_{ij}$), we obtain the power spectra 
\bea
\Delta_S^2(q)&=& -\frac{q^3}{16\pi^2{\rm Im}B(-iq)} \cr
\Delta_T^2(q)&= &-\frac{2q^3}{\pi^2{\rm Im}
A(-iq)}\;,\label{holopower}
\eea
where we have already performed the analytical continuation to Lorentzian signature through $\bar q=-iq$ and $\bar N=-iN$.

\section{Top-down model from dimensional reduction of ${\cal N}=4 $ SYM vs. $AdS_5\times S^5$}

Another holographic approach was developed in \cite{Awad:2008jf,Brandenberger:2016egn,Ferreira:2016gfg}, and we will present it in a way that can fit into the 
holographic cosmology paradigm from the previous section. 

We consider a 4+1 dimensional geometry that is a solution of the 10 dimensional type IIB equations of motion, with a metric ansatz
\be
ds^2=\frac{R^2}{z^2}[dz^2+(-dT^2+a^2(T)d\vec{x}^2)]+R^2d\Omega_5^2\;,\label{FLRWansatz}
\ee
and with a nontrivial dilaton $\phi=\phi(T)$. 

More generally, for the metric ansatz
\be
ds^2=\frac{R^2}{z^2}[dz^2+g_{\mu\nu}(x)dx^\mu dx^\nu)]+R^2d\Omega_5^2\;,
\ee
the equations of motion are 
\be
R_{\mu\nu}[g_{\rho\sigma}]=\frac{1}{2}\d_\mu\phi\d_\nu\phi\;,\;\;\;
\d_A(\sqrt{-G}G^{AB}\d_B \phi)=0\;,
\ee
where $G_{AB}$ is the 5-dimensional metric. With a flat FLRW cosmological ansatz as in (\ref{FLRWansatz}), one finds the unique solution 
\be
a(T)\propto T^{1/3}\;,\;\;\; e^{\phi(T)}=\left(\frac{T}{R}\right)^{2/\sqrt{3}}\;,\label{stiff}
\ee
which corresponds to a "stiff matter" cosmology, with equation of state $P=w\rho$, with $w=+1$. Indeed, in general for FLRW we have $a(T)\propto T^{\frac{2}{3(1+w)}}$.

Making a transformation to conformal time $t$ (usually called $\eta$), we obtain
\be
-dT^2+a^2(T)d\vec{x}^2=a^2(t)[-dt^2+d\vec{x}^2]\Rightarrow a\sim T^{1/3}\sim t^{1/2}\;,
\ee
so in particular
\be
e^{\phi(t)}=\left(\frac{t}{R}\right)^{\sqrt{3}}.
\ee

In fact, for a general homogeneous and isotropic cosmological ansatz for the metric, we have
\be
R_{00}[g_{\rho\sigma}] = -3\frac{\Ddot{a}}{a}\;,\;\;\; R_{ij}[g_{\rho\sigma}] = \left(\frac{\Ddot{a}}{a}+ 2\left(\frac{\Dot{a}}{a}\right)^2+2\frac{k}{a^2}\right)\delta_{ij}\;,
\ee
with $k=-1,0,1$ for open, flat and closed universes. So, to have a 10 dimensional solution with homogeneous dilaton, we should have
\be
\Dot{\phi}^2 = - 6\frac{\Ddot{a}}{a}\;, \;\;\;
\frac{\Ddot{a}}{a}+ 2\left(\frac{\Dot{a}}{a}\right)^2+2\frac{k}{a^2} = 0\;,
\ee
which in conformal time reads 
\be
\left(\frac{d \phi}{dt}\right)^2 = 6\left[\frac{1}{2a^4}\left(\frac{d}{dt}(a^2)\right)^2 + 2k\right]\;,\;\;\; \frac{1}{2a^2}\frac{d^2}{dt^2}(a^2)+ 2k =0. \label{eqsofm}
\ee

Solving these equations for $k=0$ gives the results before. For $k=1$ the solution is
\be
a(t)\propto |\sin(2t)|^{1/2}\;,\;\;\; e^{\phi(t)} \propto |\tan(t/R)|^{\sqrt{3}}\;,
\ee
and for $k=-1$ we have
\be
a(t)\propto |\sinh(2t)|^{1/2}\;,\;\;\; e^{\phi(t)} \propto |\tanh(t/R)|^{\sqrt{3}}.\;
\ee
We conclude that, for homogeneous dilaton, there is unique solution for each possible spatial 
``topology'' (in the restricted sense associated with the sign of the curvature, of closed, open, or flat).

Note that the original $G_{AB}$ metric was in Einstein frame, and $\phi(T)$ was the dilaton. If we make the conformal transformation
by $a(T)$ we move away from the Einstein frame. Then $\phi(T)=\phi(t)$ is the dilaton, thus $e^{\phi(t)}$ is the string coupling, corresponding in the boundary field theory to 
the YM coupling $g^2_{YM}/(4\pi)$. In terms of the time $t$ of Minkowski space, we have then a time-dependent SYM coupling,
\be
g_{YM}(t)=g_{YM,0}\left(\frac{|t|}{R}\right)^{\sqrt{3}}.
\ee

The conformal transformation on the boundary is allowed, given that the boundary field theory is conformal. However, when doing that in holography, 
we will obtain a modification of the holographic map, that will be calculated in the next subsection. 

As an aside, note that the solution
\bea
ds^2_4&=&|\sinh(2t)|\left[-dt^2+\frac{dr^2}{1+r^2}+r^2d\Omega_2^2\right]\cr
e^{\phi(t)}&=& g_s|\tanh (t/R)|^{\sqrt{3}}
\eea
is not just conformally flat, but actually asymptotically flat. For $t\rightarrow \pm \infty$, there is a coordinate transformation that takes away the conformal factor, giving $\text{AdS}_5\times S^5$ and constant dilaton in these regimes, as analyzed in \cite{Awad:2008jf}. 
However, {\em close to the strong coupling gravity region $t\sim 0$}, we get still a conformal factor deviating from 1, and the solution is the same as before, $a^2(t)\propto t$ and $e^{\phi(t)}\propto 
t^{\sqrt{3}}$.

Until now, we have presented solutions with Lorentzian signature. However, the AdS/CFT correspondence
is known to be better understood and defined in Euclidean signature space, and the Wick rotation 
to Lorentzian signature to be a difficult issue.\footnote{In particular, in many cases the 
correspondence is better 
defined in global coordinates, but there is not a clear relation between the natural Wick rotation 
in global coordinates (mapped to ``radial'' Wick rotation in the boundary Minkowski space, $r=-ir_E$) 
and the standard Wick rotation in Poincar\'{e} coordinates (mapped to the usual $t=-it_E$). Also in the case of the pp wave correspondence \cite{Berenstein:2002jq}, 
obtained as a Penrose limit of AdS/CFT, the issue of Wick rotation is extremely subtle, as found 
for instance in \cite{Berenstein:2002sa}. In all these cases, the implicit assumption is that the 
field theory dual to the Euclidean signature solution is still the Euclidean version of the 
field theory, and that is the starting point for defining the correspondence.}
Therefore, one does not simply Wick rotate the solutions we found to Euclidean signature,  but rather
considers a {\em mapping} from the above solutions to solutions of the Euclidean version of supergravity
(thus string theory), and that is where we assume that our correspondence is defined.\footnote{This
is similar to the ``domain wall/cosmology correspondence'' of Skenderis and 
Townsend \cite{Skenderis:2006fb}, used in the definition of the phenomenological holographic 
cosmology model of McFadden and  Skenderis \cite{McFadden:2009fg} and reviewed in the previous 
section. There also, one does not have a Wick rotation per se, but rather a mapping of solutions 
from ones of a domain wall type to ones of a cosmology type (similar to a double Wick rotation), 
and only then Wick rotates the domain wall from Lorentzian to Euclidean signature.}
Specifically,  the equations of motion (\ref{eqsofm}) for $k=0$ 
are invariant under the standard Wick rotation
$t=-it_E$, which means both in Lorentzian and Euclidean signature we have the same 
solutions (\ref{stiff}), and we can choose real prefactors in both cases, i.e. 
$(a(t)=a_0 (t/t_0)^{1/2}, e^{\phi(t)}=e^{\phi_0}(t/R)^{\sqrt{3}})$ and 
$(a_E(t_E)=a_{0,E}(t_E/t_{0,E})^{1/2}, e^{\phi_E(t_E)}=e^{\phi_{0,E}}(t_E/R)^{\sqrt{3}})$, 
even though the two solutions are not Wick rotations of each other (in which case we would need 
to define branch cuts, obtain complex prefactors, etc.). From now on, we will assume the Euclidean 
signature solutions and dual field theory. 

In order to embed the approach presented in this section 
into the paradigm from the last section, we need to consider how to extend it to the case when there is both a radial coordinate, and a 
time coordinate. For the general set-up of Maldacena, the wavefunction of the Universe $\psi[h_{ij}]$ is evolved in time with the Hamiltonian, which corresponds on the boundary 
to the RG flow of the correlators obtained from $Z[h_{ij}]$, as the energy scale is varied. In the framework of \cite{McFadden:2009fg}, the Wick rotation 
("domain wall/cosmology correspondence") means that time evolution is replaced by a radial "Hamiltonian" evolution, corresponding to the same, and in line with the 
usual AdS/CFT construction.

The Maldacena map is based on the fact that the wavefunction of the Universe can be thought of as a path integral, integrated over time (in the past), but 
with the boundary condition of spatial 3-metric $h_{ij}$ at the corresponding time $t$. Then it is really just a type of analytical continuation of the usual AdS/CFT
map between the partition function of the field theory, with sources $h_{ij}$, and the partition function of the gravity or string theory (written as a path integral), 
with a boundary condition of $h_{ij}$. 

But now we have both a radial direction and a time direction, and we have to decide how to generalize the set-up of Maldacena to this situation, so that maybe in a 
second step, we can relate it to the paradigm of \cite{McFadden:2009fg}.\footnote{See 
\cite{Skenderis:2008dh,Skenderis:2008dg,Christodoulou:2016nej} for early treatments of having both time and radial 
direction, though outside the holographic cosmology context we introduce here.}

There are now two possible Hamiltonians in the gravitational theory: 
both the radial one, who gives the evolution that, via the usual AdS/CFT correspondence, corresponds to the RG flow 
of the boundary field theory, and the true Hamiltonian, which gives the evolution of gravity along the time direction, and should similarly correspond to a Hamiltonian evolution 
in time in the boundary field theory. 

It seems therefore reasonable to assume that the correct prescription to use is to have, on the gravity side, a partition function with boundary condition both at time $t$ and at 
radial size $r$, which therefore is still a wavefunction of the Universe,
corresponding in field theory to a partition function integrated over time until the corresponding time $t$, and both be as usual functions of 
spatial 3-metrics $h_{ij}$,
\be
\psi[h_{ij}]_{t,r}=Z[h_{ij}]_{t,q}.
\ee

Here $q$ is the energy scale corresponding holographically to the radial direction $r$ in the bulk. The time $t$ is arbitrary,
and the path integration is assumed to be for times between $-\infty$ and $t$, but not in its future. In this way, both 
sides of the equation are functions of this time $t$, which are evolved with the Hamiltonian. Of course, in the context of 
the ``top-down'' model, the bulk will have a time-dependent Hamiltonian, which can be interpreted in terms of particle
production. The evolution with the Hamiltonian from $t$ to $t'$ 
should be equivalent with the path integration until a later time $t'$.

Next, we need to understand the effect of the integration over $t$ on both sides of the equality, and how to take it into account. 
On the gravity side, the integration over time gives the wavefunction of the 
Universe, and there is nothing we need to do with it. Since the holographic map
 is the same, the calculation of the correlators of metric fluctuations in (\ref{correl}) 
is unchanged, and we should obtain the same relation (\ref{holopower}). 

On the field theory side, we should do the path integral over the time direction until the time $t$. Because of the fact that $g_{YM}(t)$ is a positive power law, and appears in 
the denominator in the action, 
\be
e^{-S}=e^{-\int dt \frac{1}{g^2_{YM}(t)}\int d^3x {\cal L}_{\rm SYM}}\;,
\ee
the largest contribution to the weight $e^{-S}$ will be from small times. But then, if at small $t$ the fields are {\em positive}
power laws in time (which should be the case since fields must not be singular at $t=0$, and must be Taylor expandable), which would correspond to "massive KK modes" 
in a "KK" expansion in $t$ of the fields of SYM, these would give small contributions to the path integral. The leading contribution must be from the time-independent 
fields, i.e., the "KK dimensionally reduced" fields. We also should split the Lorentz indices according to this dimensional reduction, finally obtaining a 3 dimensional 
field theory action, with coupling factor integrated over time from a time $t_X\sim t_{\rm Pl}$ of the order of the Planck scale up to the relevant $t$,
\be
\int dt\frac{1}{g^2_{YM}(t)}\sim \frac{1}{g^2_{YM,0}}\int_{t_{\rm Pl}}^{t_X} \frac{dt}{(t/R)^{\sqrt{3}}}=\frac{R}{g^2_{YM,0}}\left.(t/R)^{1-\sqrt{3}}\right|_{t_{\rm Pl}}
^{t_X}\equiv \frac{RK}{g^2_{YM,0}}\equiv \frac{1}{g^2_{3d}}.
\ee
Here $K=(t_X/R)^{1-\sqrt{3}}-(t_{\rm Pl}/R)^{1-\sqrt{3}}$ is very large. 

But then the {\em effective} (dimensionles) 3 dimensional coupling is
\be
g^2_{\rm eff}\equiv \frac{g^2_{3d}N}{\bar q}=\frac{g^2_{YM,0}N}{K(R\bar q)}.
\ee

Since both $g^2_{YM,0}N\gg 1$ (from the usual holographic condition on the validity of the supergravity approximation for $AdS_5\times S^5$) 
and $K\gg 1$, we can have even $R\bar q\sim 1$, and still {\em we can choose} the effective coupling to be perturbative, $g^2_{\rm eff}<1$, though that is 
not necessary.

In this case, we see that we obtain a specific 3 dimensional field theory with generalized conformal structure, one obtained from the dimensional 
reduction of ${\cal N}=4$ SYM. However, in \cite{Afshordi:2016dvb,Afshordi:2017ihr} the best fit to the CMBR data of the {\em perturbative} phenomenological 
field theory  was analyzed, and it was found that for no fermions (introducing fermions moves the fit away from the desired region), the number of adjoint scalars
for a good match is of the order of $10^4$, which is much larger than the one obtained from dimensionally reducing ${\cal N}=4$ SYM (which is 7: 6 originally, and 
one from the $A_0$ component of the gauge field). That means that this theory does not fit the CMBR data {\em perturbatively}. 

It could be that one needs to choose a larger coupling (so as not to have $g^2_{\rm eff}<1$) in order to find the fit, though to test that we would need access to lattice data.
Or it could be that ${\cal N}=4$ SYM is just a toy model, and we would need to apply the same methods to other top down gravity dual pairs, though we will leave that 
for further work.\footnote{Note that supersymmetry itself for the gravity dual pair 
is not ruled out by the CMBR data, 
since bosons and fermions give different contributions to the fit; only a small number of fields in 3 
dimensions is, since as we said, we need of the order of $10^4$ bosons.}  
In particular, we saw that the $a(t)$ uniquely selected by the type IIB equations of motion corresponded to a "stiff matter" cosmology, with $w=1$, which 
is different than, say, inflation.

\subsection{Transformation of dilaton and operator
VEV}\footnote{This section was done in collaboration with 
Kostas Skenderis}

We could ask: where do we see the dependence on the cosmological model $a(t)$? There is not much dependence in  the constant $K$, defining $g^2_{3d}$, 
and there would be a small dependence if we took into account corrections due to non-constant field theory modes (considering "the full KK tower" of fields, 
instead of the dimensionally reduced ones only). Of course, the type IIB equations of motion only allow a specific $a(t)$, so it cannot be varied, but it still seems strange. 
Here we want to see that there is in fact one quantity that depends on it, though it should affect only correlators away from the perturbative regime. 

We have already noted that a conformal rescaling on the boundary, to go from a conformally flat space to a flat space (by the $a^2(t)$ factor that takes us from 
a cosmological model to a simple flat space), corresponds in the bulk to a coordinate transformation. 

Indeed, a conformal transformation {\em on the boundary} can be thought of as embedded in the set of general coordinate transformations 
{\em on the boundary} (conformal transformations are global $SO(4,2)$ transformations in $d=4$, embedded in the infinite dimensional "group" of 
general coordinate transformations). But by applying a conformal transformation, we just obtain a specific coordinate transformation, differing from what we have, 
which means that we can't remove the conformal factor by a conformal transformation {\em on the boundary}. 

But we can remove {\em any} conformal factor on the boundary by a coordinate transformation {\em in the bulk}, as shown in 
\cite{Skenderis:2000in}, eqs. 8,9,10.

Let us apply this procedure to our case. Writing $\rho = z^2$, the general coordinate transformation is expanded as
\bea
\rho&=&\rho' e^{-2\sigma(x')}+\sum_{k\geq 2} a_{(k)}(x')\rho'^k\cr
x^i&=& x'^i+\sum_{k\geq 1} a^i_{(k)}(x')\rho'^k\;,
\eea
which gives
\be
g'_{(0)ij}=e^{2\sigma}g_{(0)ij}
\ee
and higher orders, which don't interest us. 

For us, we have 
\be
g_{(0)ij}=a^2(t)\delta_{ij}\;,\;\;
g'_{(0)ij}=\delta_{ij}\;,
\ee
so $e^{2\sigma}=a^{-2}(t)$. That means that we only need to transform time, as
\be
t=t'+\sum_{k\geq 1}a^0_{(k)}(t') \rho'^k\;,
\ee
but not space (since the metric is space independent). Then the formulas for the relevant coefficients are (note that we are not interested in the 
transformation on $\rho$, so we don't care about $a_{(k)}$'s) 
\bea
a^0_{(1)}&=&\frac{1}{2}\d^t\sigma e^{-2\sigma}\cr
a^0_{(2)}&=&-\frac{1}{4}e^{-4\sigma}\left(\d_t \sigma g^{tt}_{(2)}+\frac{1}{2}\d^t \sigma (\d\sigma)^2+\frac{1}{2}{\Gamma^t}_{tt}\d^t\sigma \d^t\sigma\right)\;, {\rm where}\cr
g_{(2)ij}&=&\frac{1}{d-2}\left(R_{ij}-\frac{1}{2(d-1)}R g_{(0)ij}\right).
\eea
Here indices are raised and lowered with $g_{(0)ij}=a^2(t)\delta_{ij}$. 

We consider in particular the cosmological solution of the type IIB equations of motion, which has 
\be
a^2(t)=t\;,\;\;\;
e^{\phi(t)}=t^{\sqrt{3}}\Rightarrow \phi=\sqrt{3}\ln t\;,
\ee
and solves
\be
R_{ij}=\frac{1}{2}\d_i\phi \d_j\phi\;,
\ee
giving for $g_{(2)ij}$ the value
\be
g_{(2)ij}=\frac{1}{2}\left(\frac{\d_i\phi\d_j\phi}{2}-\frac{1}{6}(\d\phi)^2g_{(0)ij}\right)\;,
\ee
or more precisely
\be
g_{(2)tt}=\frac{1}{6}(\d_t\phi)^2.
\ee

We also calculate the relevant Christoffel symbol,
\be
{\Gamma^t}_{tt}=\frac{1}{2t}.
\ee

Then, after a bit of algebra, we find the coefficients
\be
a^0_{(1)}=\frac{1}{4t}\;,\;\;\;
a^0_{(2)}=\frac{1}{16t^3}.
\ee

Substituting in the coordinate transformation of the time direction, we find 
\be
t=t'+\frac{1}{4t'}\rho'+\frac{1}{16 t'^3}\rho'^2.
\ee

The scalar transformation law is $\phi'(t')=\phi(t)$, so we obtain 
\be
\phi'(t')=\phi(t)=\phi\left(t'+\frac{\rho'}{4t'}+\frac{\rho'^2}{16 t'^3}\right)=\sqrt{3}\ln \left[t'+\frac{\rho'}{4t'}+\frac{\rho'^2}{16 t'^3}\right].
\ee

Expanding near the boundary at $\rho=0$, we find 
\be
\phi'(t')=\sqrt{3}\left[\ln t'+\ln \left(1+\frac{\rho'}{4t'^2}+\frac{\rho'^2}{16 t'^4}\right)\right]
\simeq \sqrt{3}\left[\ln t'+\frac{\rho'}{4t'^2}+\frac{\rho'^2}{32 t'^4}\right].
\ee

The leading term in the $\rho'$ expansion of on-shell fields is the source on the boundary, and we see that it is unmodified in the case of $\phi(t)$. 
The second term in the expansion of $\phi(t)$ (with $\rho'$)
is related to the first, but the third (with $\rho'^2$) is related to an operator VEV on the boundary. 

That means that we have, besides the source, also an operator VEV in the ${\cal N}=4$ SYM with time dependent coupling. This coupling $g^2_{YM}(t)$ 
is unchanged, but we have obtained a nonzero VEV, of  
\be
\langle \Tr[F_{\mu\nu}^2]\rangle\propto \frac{1}{32 t'^4}\neq 0.
\ee

This operator VEV is truly dependent on the cosmological solution $a(t)$, as we have seen, and its presence should modify nonperturbatively the SYM correlators. 
But in the perturbation theory we have considered, there is no modification.\footnote{In a CFT, a state is created by a 
local operator, so a correlator in a different state is equivalent with adding  two more operators in the vacuum correlator. 
However, the perturbation theory considered here is in 3 Euclidean dimensions, 
after the ``dimensional reduction'' of the 
time direction. From this theory's point of view, we are in a nonperturbative state: at fixed time, the VEV calculated here 
is constant throughout the space, and thus is not the effect of a local operator.}

\section{Conclusions and discussion}

In this paper we have extended the holographic cosmology map $Z[h_{ij},\phi]=\psi[h_{ij},\phi]$ of Maldacena, between the wavefunction of the Universe and the 
boundary partition function, to the case where there is {\em both} an Euclidean holographic direction, and  a Minkowskian time direction, obtaining 
$\psi[h_{ij},\phi]_{t,r}=Z[h_{ij},\phi]_{t,q}$. 
Specifically, we applied this prescription to the case of a cosmological solution of the type IIB equations of motion with a time-dependent dilaton $\phi(t)$, 
where the conformal factor $a^2(t)$ relates it conformally to a flat space solution, corresponding to the usual $AdS_5\times S^5$ vs. ${\cal N}=4$ SYM in flat space.
This is therefore a "top down" holographic cosmology, obtained by a modification of the original AdS/CFT case. 
We have then proposed that to integrate over the time direction as needed, we can, in the boundary partition function, 
"dimensionally reduce" the theory on the time direction, by considering only time-independent quantities, except for the overall coupling $g_{YM}(t)$. In so doing, we obtain 
the set-up of \cite{McFadden:2009fg}, just that from a top down, as opposed to bottom up, construction.
While the resulting cosmology was not, perturbatively, consistent with the CMBR data, we could think of the possibility of either a non-perturbative match, where 
the SYM results would be obtained on the lattice, or of  using the same construction for a different top down starting point. These possibilities are left for further work. 
We have also shown that the effect of the scale factor $a(t)$ (of the cosmology) on the correlators of SYM is to introduce a nonzero 
time-dependent VEV  $\langle \Tr[F_{\mu\nu}^2]\rangle$ non-perturbatively.

\section*{Acknowledgements}
We thank Kostas Skenderis for participation at the early stages of this project, in particular for section 3.1, which 
was done together with him, and for comments on the manuscript. We also thank Robert Brandenberger for 
useful comments on the manuscript.
The work of HN is supported in part by CNPq grant 304006/2016-5 and FAPESP grant 2014/18634-9. HN would also 
like to thank the ICTP-SAIFR for their support through FAPESP grant 2016/01343-7, and to the University of Cape Town for hospitality during the final stages of this work. HB would like to thank CAPES, for supporting his work, and also McGill University for hospitality during an exchanged period when final stages of this work were completed.

\appendix

\section{Holographic calculation of the scalar and tensor two-point functions}

In this Appendix, we review the holographic calculation in \cite{Papadimitriou:2004ap,Papadimitriou:2004rz,McFadden:2010na}, relating 
$\langle \delta h_{ij} \delta h_{kl} \rangle$ correlators (experimentally derived from the CMBR) to $\langle T_{ij} T_{kl} \rangle$ 
correlators in the ${\cal N}=4$ SYM field theory, using the radial Hamiltonian 
formalism. 

Consider an asymptotically AdS metric in Fefferman-Graham coordinates,
\be
ds^2=\frac{1}{z^2}\left[dz^2+\left(g_{(0)ij}+...+z^dg_{(d)ij}+...\right)\right]\;,\label{above}
\ee

The one-point function of the energy-momentum tensor in the presence of sources is then 
\be
\langle T_{ij}(x)\rangle =-\frac{1}{\sqrt{g_{(0)(x)}}}\frac{\delta W[g_{(0)},...]}{\delta g^{ij}_{(0)}(x)}\;,
\ee
where $W$, the generating functional of connected graphs, equals by the AdS/CFT prescription (minus) the on-shell action $S_{\rm on-shell}$.

We use a radial Hamiltonian formulation for AdS gravity, with $r$, 
\be
z=e^{-r}\;,
\ee
acting as "time" in the "ADM parametrization"
\be
ds^2=g_{\mu\nu}dx^\mu dx^\nu=\hat \gamma_{ij}d\hat x^i d\hat x^j+2N_i d\hat x^i dr +(N^2+N_iN^i)dr^2.
\ee

Then the asymptotically AdS metric is
\be
ds^2=dr^2+g_{ij}(r,x)dx^idx^j
\ee
and $g_{(p)ij}$ means (as in (\ref{above})) the expansion of $g_{ij}$ in $z^{2p-2}=e^{(2-2p)r}$. 

Then, like in the usual ADM construction, we can always choose a gauge such that $N=1, N_i=0$, and the ADM parametrization becomes the same as 
the Fefferman-Graham expansion above, with 
\be
\hat \gamma_{ij}=g_{ij}=\frac{1}{z^2}(g_{(0)ij}(x)+{\cal O}(z^2))\simeq
e^{2r}g_{(0)ij}(x).
\ee

The resulting on-shell action is 
\be
S_{\rm on-shell}=-\frac{1}{8\pi G_N}\int_{r_0}^{r_\epsilon} dr\int d^dx\sqrt{\hat\gamma}N\left[\hat R+8\pi G_N (\tilde T_{ij}-{\cal L}_m)\right]\;,
\ee
and one defines the canonically conjugate momentum to $\hat\gamma_{ij}$ as (at the position $r_\epsilon=1/\epsilon$, close to the boundary
at $z=0$)
\be
\pi^{ij}(r_\epsilon,x)=\frac{\delta S_{\rm on-shell}}{\delta \hat \gamma_{ij}(r_\epsilon,x)}.
\ee

We obtain 
\bea
\d_r&\simeq& \int d^dx \; 2\hat \gamma_{ij}\frac{\delta}{\delta\hat\gamma_{ij}}+\int d^dx (\Delta_I-d)\Phi_I\frac{\delta }{\delta \Phi^I}\cr
&=&\delta_D(1+
{\cal O}(e^{-2r}))\;,
\eea
where $D$ is the dilatation operator. 

Thus we can identify the radial expansion with the expansion in the eigenfunctions of the dilation operator. 
In particular, we could do that for the canonical momentum, which is found in the radial picture to equal
\be
\pi_{ij}=\frac{\sqrt{g}}{16\pi G_N}(K_{ij}-K\hat\gamma_{ij})\;,
\ee
where $K_{ij}$ is the extrinsic curvature of the radial surface, 
\be
K_{ij}=\frac{1}{2}\d_r g_{ij}\rightarrow \frac{1}{2}\delta_D g_{ij}\;,
\ee
$K=K_{ij}\hat \gamma^{ij}$,
and expand the canonical momentum in eigenvalues of $\delta_D$,
\be
\delta_D\pi_{ij}^{(n)}=-n\pi_{ij}^{(n)}.
\ee
This would not be important in the unrenormalized case, but in the renormalized case, it is. 

Then, identifying $S_{\rm on-shell}$ with $-W$ as before, we obtain a relation between the one-point function of the energy-momentum 
tensor and the canonical momentum conjugate to $\hat \gamma_{ij}$,
\be
\langle T_{ij}\rangle=-\frac{2}{\sqrt{g}}\pi_{ij}\;,
\ee
which is valid even in the renormalized case, provided we keep the piece of engineering dimension equal to the spatial one, $d$ (3 in the 
physical case), so 
\bea
\langle T_{ij}\rangle &=&\left(-\frac{2}{\sqrt{g}}\pi_{ij}\right)_{(d)}=-\frac{1}{8\pi G_N}(K_{ij}-K\hat\gamma_{ij})_{(d)}
=-\frac{1}{8\pi G_N}(K_{(d)ij}-K_{(d)}\hat \gamma_{ij})\cr
&=&-\frac{1}{16\pi G_N}(\d_r g_{(d)ij}-\hat \gamma^{kl}\d_r g_{(d)kl}\hat \gamma_{ij})\cr
&\simeq & -\frac{d}{16\pi G_N}g_{(d)ij}.
\eea

The 2-point function is found from the variation of the one-point function in the presence of sources,
\be
\delta \langle T_{ij}(x)\rangle=-\int d^3y \sqrt{g_{(0)}}\left(\frac{1}{2}\langle T_{ij}(x)T_{kl}(y)\delta g_{(0}^{kl}(y)+{\cal O}(\delta \phi_I)\right)\;,
\ee
so that
\be
\langle T_{ij}(x)T_{kl}(y)\rangle =\frac{1}{\sqrt{g_{(0)}}}\frac{\delta}{\delta g_{kl}^{(0)}(y)}\langle T_{ij}(x)\rangle
=\frac{1}{\sqrt{g_{(0)}}}\frac{\delta}{\delta g_{kl}^{(0)}(y)}\left(-\frac{2}{\sqrt{g}}\pi_{ij}\right)_{(d)}.
\ee

The right-hand side, when we take out the trivial index structure, was in a sense the definition of the linear response functions, 
which to linear order satisfy
\be
E=\frac{\delta \pi_q^\gamma}{\delta \gamma_q}+{\rm nonlinear}\;,\;\;\;
\Omega=\frac{\delta \pi^\zeta}{\delta \zeta_q}+{\rm nonlinear}\;,
\ee
so after decomposing, in momentum space
\be
\langle T_{ij}(q)T_{kl}(-q)\rangle =A(q)\pi_{ijkl}+B(q)\pi_{ij}\pi_{kl}\;,
\ee
we find 
\be
A(q)=4E_{(0)}(q)\;,\;\;\;
B(q)=\frac{1}{4}\Omega_{(0)}(q).
\ee

\bibliographystyle{utphys}
\bibliography{HoloCosmoPaper2}


\end{document}